\documentclass[%
 reprint,
nofootinbib,
 amsmath,amssymb,
 aps,
]{revtex4-2}

\usepackage{hyperref}

\usepackage{graphicx}% Include figure files
\usepackage{dcolumn}% Align table columns on decimal point
\usepackage{bm}% bold math

\usepackage[mathlines]{lineno}% Enable numbering of text and display math
\begin{document}

%\preprint{APS/123-QED}

\title{Revising Indirect Dark Matter Constraints with Updated Astrophysical $J$-Factor Priors}

\author{Giacomo D'Amico}
\affiliation{Instituto de Física de Altas Energías (IFAE), Barcelona, Spain}

%\date{\today}% It is always \today, today,
             %  but any date may be explicitly specified

\begin{abstract}
Indirect searches for particle dark matter with gamma-ray experiments have
produced a large number of constraints on the annihilation cross section (or decay lifetime) over a
wide range of dark matter masses.
These constraints depend critically on the assumed astrophysical $J$ factor and its uncertainty,
which encodes the dark matter distribution in the target and  represents
the dominant source of systematic uncertainty.
As improved observational data and dynamical modeling are expected to revise
current $J$-factor determinations, many published limits risk becoming obsolete
unless the full experimental analyses are repeated.
In this work we present a general and statistically consistent framework for  updating published dark matter limits when revised $J$-factor estimates become
available, without requiring access to the full experimental likelihood.
 We derive an analytical expression that quantifies the impact of astrophysical uncertainties on dark matter limits, treating both Gaussian and log-normal priors on the $J$ factor.
The formalism is validated through toy Monte Carlo simulations, including dedicated studies of its numerical stability under successive reinterpretations, and demonstrate their
accuracy by reproducing published limits.
Lastly, we further show that the formalism naturally extends to the combination of
multiple targets through a simple numerical procedure, allowing limits to be combined and
updated using only publicly available information.
The proposed method is intended as a complementary reinterpretation tool for situations in which a complete experimental reanalysis is impractical, offering a practical means to preserve and extend the scientific
relevance of published dark matter constraints across present and future
experiments.

\end{abstract}
%\keywords{Suggested keywords}%Use showkeys class option if keyword
                              %display desired
\maketitle

\section{Introduction}
\label{sec:introduction}

Indirect searches for particle dark matter aim at detecting the products of dark
matter annihilation or decay in astrophysical environments, with gamma rays
playing a central role due to their ability to propagate unattenuated from the
source to the observer.
Over the past decade, major ground-based and space-based experiments have
produced increasingly stringent upper limits on the dark matter annihilation
cross section \cite{doro2021fundamental}, exploiting observations of targets with high dark-matter
densities and low astrophysical backgrounds.
Notable examples include results from the MAGIC, H.E.S.S., VERITAS, and
\textit{Fermi}-LAT collaborations, as well as more recent constraints from
wide-field instruments such as LHAASO
\cite{Fermi-LAT:2015att,MAGIC:2021mog,abramowski2011search,archambault2017dark,LHAASO:2024upb}.

For many of these searches, dwarf spheroidal galaxies of the Local Group provide some of the most sensitive probes, owing to their large mass-to-light ratios and their relative proximity \cite{battaglia2022stellar}. More generally, for any astrophysical target used in indirect dark matter searches, the expected gamma-ray flux from dark matter annihilation factorizes into a particle-physics term and an astrophysical term, the so-called $J$ factor, which encodes the line-of-sight integral of the squared dark matter density. The latter depends on the dark matter distribution within the target and is therefore subject to astrophysical uncertainties~\footnote{In this work we focus on the case of dark matter annihilation, for which
the astrophysical factor depends on the square of the dark matter density.
The case of decaying dark matter, where the corresponding astrophysical factor
depends linearly on the density, can be treated as a straightforward
reformulation of the annihilation case and will be briefly discussed at the end
of the paper.}.
As a consequence, the inferred upper limits on the dark matter annihilation cross section are not determined solely by the observed gamma-ray data, but also depend on the astrophysical $J$ factor, which enters the likelihood as an externally constrained nuisance parameter.

This structure, in which the dark matter annihilation cross section (the parameter of interest) enters multiplicatively with the astrophysical $J$ factor (an externally constrained nuisance parameter), is common in astroparticle physics and beyond. In such cases, the published upper limits on the annihilation cross section are intrinsically tied to the prior assumptions adopted for the $J$ factor at the time of the analysis. When improved determinations of the $J$ factor become available, the statistical interpretation of those published upper limits may change, even though the underlying gamma-ray data remain unchanged.

In the context of indirect dark matter searches, uncertainties on the $J$ factor represent the dominant
systematic effect.
They originate from the limited number of stellar tracers, from modeling
assumptions in the dynamical analysis, and from degeneracies between the dark
matter density profile and the stellar velocity anisotropy
\cite{walker2007velocity,martinez2015robust,strigari2018dark}.
Importantly, these uncertainties are not static.
Significant progress is expected  from a broad range of ongoing
and upcoming observational efforts aimed at improving both the census and the
kinematic characterization of dwarf spheroidal galaxies.
Wide-field photometric surveys such as DES~\cite{sanchez2016dark}, Pan-STARRS~\cite{flewelling2020pan}, HSC-SSP~\cite{aihara2018hyper}, and future
facilities including Euclid~\cite{mellier2024euclid} and ARRAKIHS~\cite{van2024arrakihs} are expected to significantly increase
the number of known faint and ultrafaint systems and to provide improved
constraints on their structural properties
\cite{dark2016dark,aihara2022third}.
At the same time, large spectroscopic programs such as SDSS~\cite{york2000sloan}, DESI~\cite{dey2019overview}, WEAVE~\cite{jin2024wide}, and
related follow-up campaigns will deliver increasingly precise stellar velocity
measurements, which are essential to infer the underlying gravitational
potential through Jeans analyses
\cite{fan2000dwarfs}.
In parallel, the \textit{Gaia} \cite{prusti2016gaia} mission has already had a transformative impact on
the study of nearby dwarf galaxies by providing high-precision astrometric data,
enabling improved membership selection and proper-motion measurements, with
further refinements expected from its final data releases
\cite{brown2018gaia,gaia2023gaia}.
Together with methodological advances in Jeans modeling and alternative
phase-space approaches
\cite{read2019dark,hayashi2020diversity,errani2023dark},
these developments are expected to substantially improve the accuracy and
robustness of $J$-factor determinations.

As a result, current estimates of the astrophysical $J$ factors and their
associated uncertainties are likely to be revised in the coming years,
potentially leading to non-negligible shifts in the inferred dark matter
constraints derived from indirect detection experiments.
This situation poses a practical challenge: published upper limits on the dark
matter annihilation cross section are intrinsically tied to the specific
astrophysical priors adopted at the time of the analysis.
Once improved determinations of the $J$ factors become available, these limits
may become outdated, unless the full experimental likelihood is reanalyzed.
Such reanalyses are often impractical, as they require access to proprietary data,
detailed instrument response functions, and complex analysis pipelines.
Complementary to the recasting framework introduced in Ref.~\cite{d2025recasting}, where upper limits on dark matter annihilation and decay were reinterpreted for alternative particle-physics models without access to the full experimental likelihood, the goal of this work is to provide a framework to update published upper limits when revised $J$-factor determinations become available, likewise without requiring access to the full experimental likelihood.
Building on a quadratic approximation of the likelihood and an explicit treatment
of astrophysical uncertainties, we derive analytical expressions that quantify
the impact of $J$-factor uncertainties on the inferred limits for single targets,
and we show how the same framework can be extended to the combination of multiple
targets through a simple numerical procedure. 
This approach allows existing constraints to remain scientifically useful and
directly comparable as astrophysical knowledge improves~\footnote{It is worth noting that reinterpreting constraints with updated $J$-factor uncertainties also requires assuming the same observed data as in the original analysis. The observed data generally depend on the angular integration radius $\theta$, which defines the region of interest used to extract the gamma-ray data. Consequently, the integration angle must be kept fixed to the value $\theta^*$ adopted in the original analysis, such that the data remain unchanged and the $J$-factor uncertainty is fully encoded in the probability distribution of the integrated quantity $J(<\theta^*)$.}.
In this sense, the method presented here is complementary to experimental
analyses, providing a lightweight and transparent tool to reinterpret published
limits in light of future progress in the determination of dark matter
distributions in astrophysical systems. The proposed prescription is not intended to replace a full experimental reanalysis whenever the original likelihood is available, but rather to provide a statistically motivated approximation in situations where only the published upper limit and the corresponding astrophysical prior are accessible.

Although motivated by $J$-factor uncertainties in dark matter searches, the formalism developed here applies more broadly to any inference problem in which the predicted signal depends multiplicatively on externally constrained nuisance parameters. As experimental data increasingly outlive specific modeling assumptions, such reinterpretation tools become essential for preserving the long-term scientific value of published results.

The paper is organized as follows.
In Sec.~\ref{sec:dm_signal_model} we briefly review the gamma-ray signal from annihilating dark matter and
introduce the notation used throughout the paper.
In Sec.~\ref{sec:quadartic_lkl} we discuss the quadratic approximation of the likelihood and the
inclusion of astrophysical priors.
In Sec.~\ref{sec:Penalty} we derive analytical penalty factors for Gaussian and log-normal
$J$-factor priors.
Section~\ref{sec:mc_validation} validates the analytical formalism using toy Monte Carlo simulations and investigates the stability of the proposed prescription under successive reinterpretations.
In Sec.~\ref{sec:recast_realdata} we demonstrate the applicability of the method by reproducing published
limits that explicitly include astrophysical uncertainties.
Finally, in Sec.~\ref{sec:combined_likelihood} we discuss the extension of the framework to the combination
of multiple targets and validate the numerical procedure against published
combined limits.

% -------------------------------------------------
\section{Gamma-ray signal from annihilating dark matter}
\label{sec:dm_signal_model}

Indirect searches for particle dark matter look for gamma rays produced by
annihilation of a DM particle $\chi$ in astrophysical targets. For a target
observed over a solid angle $\Delta\Omega$, the expected differential flux can
be written in factorized form as
\begin{equation}
\frac{d\Phi}{dE}(E)=
J \;
\left(
\frac{\langle\sigma v\rangle}{8\pi\,k\,m_\chi^2}\,
\frac{dN_\gamma}{dE}
\right),
\label{eq:dm_flux_th_ann_short}
\end{equation}
where the particle-physics term depends on the DM mass $m_\chi$, the velocity-averaged
annihilation cross section $\langle\sigma v\rangle$, and the gamma-ray yield per
annihilation $dN_\gamma/dE$ for the chosen final state(s). The factor $k$ accounts
for the DM nature ($k=1$ for Majorana, $k=2$ for Dirac).

All astrophysical information is contained in the $J$ factor,
\begin{equation}
J \equiv \int_{\Delta\Omega} d\Omega \int_{\rm l.o.s.} dl\;
\rho_\chi^2(l,\Omega),
\label{eq:j_ann_short}
\end{equation}
the line-of-sight integral of the squared DM density $\rho_\chi$, integrated over
the region of interest. For targets such as dwarf spheroidal galaxies, the
uncertainty on $J$ inferred from stellar kinematics is often a leading systematic
in the final constraints.

From the experimental side, the predicted number of signal events can be expressed as a linear function of $\langle\sigma v\rangle$,
\begin{align}
s &= \langle\sigma v\rangle \; J \;  K,
\label{eq:si_short}
\end{align}
where $K$ collects the exposure, instrument response,
and spectral and spatial integrals (for more detailed formulations, see, e.g., Ref.~\cite{rico2020gamma}).

No unambiguous gamma-ray detection of dark matter has been established to date, so current analyses typically assume compatibility with the background-only hypothesis\footnote{This assumption underlies the analytical treatment developed in this work. In the presence of a non-negligible excess, the reconstructed limits should instead be regarded as an approximate reinterpretation, whose accuracy depends on the degree to which the best-fit signal departs from the null hypothesis. As discussed in Sec.\ref{sec:mc_validation}, this approximation has been quantitatively validated through toy Monte Carlo simulations, including dedicated tests of its stability under successive reinterpretations.} and set upper limits on $\langle\sigma v\rangle$ as a function of $m_\chi$ (and of the assumed annihilation channel). 

These limits are determined by two ingredients: (i) the signal model, which relates the annihilation cross section to the expected number of signal counts through Eq.~\eqref{eq:si_short}, and (ii) the observed gamma-ray data (typically expressed as counts in ON and OFF regions), with the astrophysical normalization $J$ treated as a nuisance parameter.

\section{Quadratic approximation of the likelihood}
\label{sec:quadartic_lkl}

To derive an upper bound on the dark matter annihilation cross section,
$\langle\sigma v\rangle^{\rm UL}$, experimental searches usually rely on a
binned likelihood framework that compares the predicted signal contribution
to the observed gamma-ray data across reconstructed energy intervals
(see, e.g., Refs.~\cite{rico2020gamma,abramowski2011search,lefranc2016dark}.
We denote, for simplicity, the parameter of interest  $\langle\sigma v\rangle$ by $x$.
We assume from the outset that the analysis depends on an astrophysical $J$-factor, treated as a nuisance parameter constrained by an external measurement~\footnote{This assumption applies to analyses in which the spatial morphology of the dark matter signal is kept fixed and updates to the astrophysical model modify only the overall normalization of the signal. Cases in which revised astrophysical determinations also alter the spatial signal template lie outside the scope of the present work.}.

The total (binned) log-likelihood can  be expressed as a sum of bin-wise (typically energy bins) contributions $f_i$ 
\cite{DAmico:2022psx,conrad2015statistical},
\begin{equation}
-2\ln \mathcal{L} = -2 \sum_i \ln  \mathcal{L}_i\!\left(s_i\mid D_i\right)
\equiv 2\sum_i f_i\!\left(s_i\right),
\label{eq:def_fi}
\end{equation}
where $D_i$ denotes the observed data in bin $i$ (e.g.\ counts) and $s_i$ denotes the expected signal contribution in bin $i$.

For annihilation signals, the expected counts scale linearly with the product $xJ$,
and we write
\begin{equation}
s_i(x,J) = K_i\,J\,x \equiv K_i z,
\label{eq:si_linear}
\end{equation}
where $K_i$ collects all bin-dependent factors and for conveniece we have defined the composite variable $z \equiv J\,x $.

Let $\hat z$ denote the value of $z$ that minimizes $-2\ln\mathcal{L}(z)$.

Defining $f_i''(s)$ as the second derivatives of $f_i(s)$
with respect to $s$, a Taylor expansion of the log-likelihood around $z=\hat z$
yields, to second order \cite{wald1943tests},
\begin{equation}
-2\ln \mathcal{L}(z)
\simeq
\sum_i \left(\frac{\partial s_i}{\partial z}\right)^2 \cdot
f_i''\cdot (z-\hat z)^2 .
\end{equation}
Since $\partial s_i/\partial z = K_i$, this becomes
\begin{equation}
-2\ln \mathcal{L}(z)
\simeq
\left(\sum_i K_i^2 f_i''\right)\,(z-\hat z)^2 .
\end{equation}
Introducing the shorthand
\begin{equation}
A \equiv \sum_i K_i^2 f_i'',
\end{equation}
the quadratic approximation can be written compactly as
\begin{equation}
-2\ln \mathcal{L}(x,J)
\simeq
A\,(z-\hat z)^2
=
A\,(Jx-\hat z)^2,
\label{eq:general_quadratic_lkl}
\end{equation}
where $A$ should be understood as the local curvature of the full likelihood
with respect to the signal-normalization variable $z=Jx$, evaluated around
the best-fit point. It therefore encodes, in an effective way, the information from the exposure, instrumental response functions, binning, and spectral and spatial modeling adopted in the original analysis.

%Finally  

\subsection{Including the prior on the \texorpdfstring{$J$}{J}-factor}

As shown in Sec \ref{sec:dm_signal_model}, the expected gamma-ray signal from dark matter annihilation depends on the
astrophysical $J$ factor. Since $J$ is not determined by the gamma-ray data themselves but is
inferred from independent stellar-kinematic measurements, it must be treated as
a nuisance parameter constrained by an external prior.

Within a likelihood-based framework, this information is incorporated by
including in the likelihood a prior probability density $\pi$ for $J$:
\begin{equation}
-2 \ln \mathcal{L}(x,J)
\xrightarrow{} 
-2 \ln \mathcal{L}(x,J) \, -2 \ln \pi(J).
\label{eq:general_lkl_w_prior}
\end{equation}

In the frequentist approach, limits on $x$ are obtained by profiling over the
nuisance parameter $J$. The specific functional form of
$\pi(J)$ depends on how uncertainties on the $J$ factor are modeled. Two choices will be considered in the next sections: a Gaussian prior on $J$ itself, and a
log-normal prior corresponding to a Gaussian distribution in $\ln J$.

\subsubsection{Gaussian prior}\label{sec:Gaussian_prior}

We assume a Gaussian prior for the astrophysical $J$-factor with mean $J_0$
and standard deviation $\sigma_J$. Up to an irrelevant additive constant, this prior contributes to the log-likelihood as
\begin{equation}
-2\ln \pi(J)
=
\frac{(J-J_0)^2}{\sigma_J^2},
\label{eq:prior_term}
\end{equation}

which when substituted in Eq. \eqref{eq:general_lkl_w_prior} gives
\begin{equation}
-2\ln \mathcal{L}(x,J) \simeq
A\,(x J-\hat z)^2 +  B (J-J_0)^2,
\label{eq:final_form_sigmaJ}
\end{equation}
where we recall that $z \equiv Jx$ and for convenience we used $B \equiv 1/\sigma_J^2 $.

At this point we profile over the nuisance parameter $J$ by minimizing
$-2\ln \mathcal{L}(x,J)$ with respect to $J$ at fixed $x$.
Taking the derivative with respect to $J$ and setting it to zero yields
\begin{equation}
\frac{\partial}{\partial J}
\left[
A\,(Jx-\hat z)^2 + B\,(J-J_0)^2
\right]
= 0 .
\end{equation}
Evaluating the derivative explicitly and solving for $J$, we obtain
\begin{equation}
J(x)
=
\frac{B\,J_0 + A\,x\,\hat z}{B + A\,x^2}.
\label{eq:J_profile}
\end{equation}

Therefor, one finds
\begin{equation}
-2\ln \mathcal{L}(x)
= -2\ln \mathcal{L} \!\left(x, J(x)\right)
\simeq
A\,B\ \frac{(J_0 \; x - \hat z )^2}{B + A\,x^2}.
\label{eq:profiled_ll}
\end{equation}

\subsubsection{Log-normal prior}

In many indirect-detection analyses, the astrophysical $J$-factor is modeled with
a \emph{log-normal} uncertainty~\cite{CTAO-dphs}. Equivalently, one assumes that
\begin{equation}
y \equiv \ln J
\end{equation}
is Gaussian distributed.  So that, up to an irrelevant normalization constant,
\begin{equation}
-2\ln \pi(y) = \frac{(y-y_0)^2}{\sigma^2}.
\end{equation}
As done in  for Eq. \eqref{eq:final_form_sigmaJ}, the log-likelihood is, up to an additive constant,
\begin{equation}
-2\ln \mathcal{L}(x,y)
\simeq
A\,(e^{y}x-\hat z)^2 + B\,(y-y_0)^2.
\label{eq:lognormal_start}
\end{equation}
where again for convenience we defined $B \equiv 1/\sigma^2 $.

To profile over the nuisance parameter $y$ one solves
\begin{equation}
\frac{\partial}{\partial y}
\left[
A\,(e^{y}x-\hat z)^2 + B\,(y-y_0)^2
\right]=0.
\end{equation}
Expanding the derivative gives
\begin{equation}
A e^{2y}x^2 - A e^{y}x\,\hat z = -B\,(y-y_0).
\label{eq:profile_eq_general}
\end{equation}
The profiled value $y(x)$ is defined implicitly by Eq.~\eqref{eq:profile_eq_general}
and yields the profile likelihood
\begin{equation}
-2\ln \mathcal{L}(x)
=
-2\ln \mathcal{L}(x,y(x)).
\end{equation}

\section{Upper limit and the \texorpdfstring{$J$}{J}-factor Penalty}\label{sec:Penalty}
The UL on x, \(x_{UL}\), can be obtained by setting 
\begin{equation}
-2\ln \mathcal{L}(x_{UL})=\lambda,
\label{eq:lambda_def}
\end{equation}
where \(\lambda\) determines the confidence level (CL) used for the UL. For instance, a one-sided \(95\%\) confidence level  UL is obtained with \(\lambda = 2.71\)~\cite{Rolke:2004mj}.

Since no statistically significant excess is observed in current searches, the likelihood is maximized at vanishing signal strength, implying $\hat{z} =0$, and from Eqs.~\eqref{eq:lambda_def}  and ~\eqref{eq:general_quadratic_lkl},  one expects the upper limit to scale as
\begin{equation}
x_{\rm UL} \simeq   
\frac{\sqrt{\lambda}}{\sqrt{A}}\,
\frac{1}{J_0}\,
\mathcal{P}(\sigma_J),
\label{eq:generic_ul_form}
\end{equation}
where   $A$ encodes the curvature of the log-likelihood\footnote{An explicit analytical expression for $A$ in terms of instrumental and spectral quantities has been derived in
Ref.~\cite{d2025recasting}.} and 
$J_0$ is the central value of the Gaussian or Log-Normal prior on the astrophysical $J$ factor.
$\mathcal{P}(\sigma_J)$ is a penalty that accounts for the uncertainty on
$J$, satisfying the following condition,
\begin{equation}
    \lim_{\sigma_J\rightarrow 0} \mathcal{P}(\sigma_J) = 1,
\end{equation}
which ensure that in the absence of uncertainty on the $J$-factor ($\sigma_J\to 0$),
Eq.~\eqref{eq:generic_ul_form} reduces to
\begin{equation}
x_{\rm UL} = \frac{\sqrt{\lambda/A}}{J_0},
\label{eq:UL_limit_sigma_zero} 
\end{equation}
recovering the standard result in case of fixed $J$-factor. (see Ref. \cite{d2025recasting}).

The purpose of this section is to derive an explicit analytical form for
$\mathcal{P}(\sigma_J)$ under different assumptions on the prior for the $J$-factor. Such an explicit analytical form would provide an immediate recipe to update a published
upper limit on the annihilation cross section when an improved determination of
the $J$-factor becomes available.
Consider an analysis that reported an upper limit on $x\equiv\langle\sigma v\rangle$
under a given $J$-factor prior characterized by $(J_{\rm old},\sigma_{\rm old})$.
If a new astrophysical analysis yields an updated prior
$(J_{\rm new},\sigma_{\rm new})$, and assuming that the experimental likelihood and the threshold $\lambda$ are unchanged, 
 from Eq.~\eqref{eq:xul_null} the updated limit can be written as
\begin{equation}
\langle\sigma v\rangle_{\rm UL}^{\rm new}
=
\langle\sigma v\rangle_{\rm UL}^{\rm old}\,
 \frac{J_{\rm old}}{J_{\rm new}} \frac{\mathcal{P}(\sigma_{\rm new})}
{\mathcal{P}(\sigma_{\rm old})}.
\label{eq:ul_update_rule}
\end{equation}

This update rule reduces to the familiar scaling
$\langle\sigma v\rangle_{\rm UL}\propto 1/J$ in the limit of negligible
$J$-factor uncertainty ($\sigma_J\to 0$), and generalizes it by accounting for
the penalty induced by the astrophysical uncertainty.

\subsection{Penalty for Gaussian prior}

Solving Eq.  \eqref{eq:lambda_def} and using the quadratic approximation  in Eq. \eqref{eq:profiled_ll} for a log-likelihood with Gaussian prior on $J$ , we have  two roots for $x$,
\begin{equation}
x_{\pm}
=
\frac{A B J_0 \hat z \;\pm\;
\sqrt{A B \lambda\left(BJ_0^2-\lambda + A\hat z^{\,2}\right)}}
{A B J_0^2 - A\lambda}
\label{eq:x_roots}
\end{equation}
The upper limit $x_{\rm UL}$ is then chosen as the physically relevant solution, i.e. the one with the positive sign~\footnote{
In the limit of vanishing $J$-factor uncertainty ($\sigma_J\to 0$, or $B\to\infty$),
the prior enforces $J=J_0$ exactly and the likelihood depends only on
$z=J_0 x$. The upper limit then reduces to
$x_{\rm UL}=(\sqrt{\lambda/A}+\hat z)/J_0$, recovering the standard result that,
at the likelihood minimum ($\lambda=0$), $x=\hat z/J_0$.} (the minus sign solution correspond to the lower upper limit).

Inserting $\hat z=0$ into Eq.~\eqref{eq:x_roots}, using  $B=1/\sigma_J^2$, and selecting the positive-sign
solution, the upper limit on $x$ becomes
\begin{equation}
x_{\rm UL}
=
\sqrt{\frac{\lambda}{A}}\;
\frac{1}{\sqrt{J_0^2 - \lambda\,\sigma_J^2}}
=
\sqrt{\frac{\lambda}{A}} \frac{1}{J_0} \frac{1}{\sqrt{1 - \lambda\,r^2}}
,
\label{eq:xul_null}
\end{equation}
where $r= \sigma_J/J_0$ is the relative uncertainty on the J-factor.

Comparing Eqs. \eqref{eq:generic_ul_form} and \eqref{eq:xul_null}, we obtain that for a Gaussian prior on the astrophysical $J$-factor, the penalty is 
\begin{equation}
   \mathcal{P}(r) = \frac{1}{ \sqrt{1-\lambda r^2} }, \qquad r\equiv \sigma_J/J_0.
   \label{eq:gaussian_penalty}
\end{equation}

which quantifies the degradation of the upper limit due to astrophysical
uncertainty: the farther one moves from the minimum of the log-likelihood
(i.e.\ the larger $\lambda$ is), or the larger the relative uncertainty $r$,
the stronger the penalty on the inferred constraint on $x$.

The validity of Eq.~\eqref{eq:gaussian_penalty} requires
\begin{equation}
J_0^2 > \lambda\,\sigma_J^2 ,
\end{equation}
or equivalently that the relative uncertainty on the $J$-factor is not too large.
If this condition is violated, the Gaussian prior becomes effectively flat over
the region probed by the likelihood, as the $J$-factor must be, by definition, non-negative, and the quadratic approximation underlying
the profile-likelihood construction breaks down.
In that regime, the experiment loses constraining power on $x$, reflecting the
fact that an unconstrained astrophysical normalization prevents a meaningful
upper limit on the annihilation cross section.

\subsection{Penalty for Log-normal prior}

Under the null-hypothesis assumption of no
dark-matter signal, for which the best-fit  signal normalization $\hat{z}$ vanishes, the profiling condition \eqref{eq:profile_eq_general} simplifies to
\begin{equation}
A e^{2y(x)}x^2 = -B\,(y(x)-y_0).
\label{eq:profile_eq_null}
\end{equation}
Evaluating the likelihood \eqref{eq:lognormal_start} at the profiled $y(x)$ and using
Eq.~\eqref{eq:profile_eq_null} to eliminate the $Ae^{2y}x^2$ term, one obtains
\begin{align}
-2\ln \mathcal{L}(x)
\simeq &
-B\,(y(x)-y_0) + B\,(y(x)-y_0)^2
=  \nonumber
\\ 
& = B\,(y(x)-y_0)\,(y(x)-y_0-1). 
\label{eq:profile_ll_null}
\end{align}

Let us define
\begin{equation}
\varepsilon \equiv y_0 -y.
\end{equation}
Then Eqs.~\eqref{eq:profile_ll_null} and \eqref{eq:lambda_def} imply
\begin{equation}
B\,\varepsilon(\varepsilon+1)=\lambda,
\end{equation}
whose solutions are
\begin{equation}
\varepsilon_\pm
=
\frac{1}{2}\left(\pm \sqrt{1+\frac{4\lambda}{B}} - 1 \right)
=
\frac{1}{2}\left(\pm \sqrt{1+4\lambda\sigma^2} - 1 \right).
\label{eq:epsilon_pm}
\end{equation}

From Eq.~\eqref{eq:profile_eq_null} we require $-(y-y_0)=\varepsilon>0$. This selects
\begin{equation}
\varepsilon \equiv \varepsilon_+ = \frac{1}{2}\left(\sqrt{1+4\lambda\sigma^2} -1   \right) \geq   0.
\label{eq:epsilon_choice}
\end{equation}

Finally, solving Eq.~\eqref{eq:profile_eq_null} for $x$ gives
\begin{equation}
x_{\rm UL}
=
e^{-y(x)}\sqrt{\frac{B}{A}}\,\sqrt{y_0-y(x)}
=
\frac{1}{\sqrt{A} } \frac{1}{J_0} \frac{\sqrt{\varepsilon}}{\sigma} \;e^{\varepsilon}.
\label{eq:xul_lognormal_general}
\end{equation}

Comparing Eqs. \eqref{eq:generic_ul_form} and \eqref{eq:xul_lognormal_general}, we obtain that for a Log-Normal prior on the astrophysical $J$-factor, the penalty is 
\begin{equation}
\mathcal{P}(\sigma)\equiv \frac{\sqrt{\varepsilon}}{\sqrt{\lambda} \sigma}\,e^{\varepsilon}, \qquad \varepsilon = \sqrt{\frac{1}{4} + \lambda \sigma^2} - \frac{1}{2}
\label{eq:penalty_lognormal}
\end{equation}

In the limit of negligible uncertainty on the $J$-factor ($\sigma\to 0$),
one has 
\begin{equation}
\varepsilon = \lambda \sigma^2 +\mathcal{O}(\sigma^{4}),
\qquad
e^{\varepsilon}=1+\mathcal{O}(\sigma^2).
\end{equation}
and therefore
\begin{equation}
\mathcal{P}(\sigma)\to \frac{ \sqrt{\lambda}}{\sqrt{\lambda}}  = 1,
\end{equation}
recovering  $x_{\rm UL}=\sqrt{\lambda/A}/J_0$ which is the result when $J$ is known exactly (the same result was obtained in Eq. \eqref{eq:UL_limit_sigma_zero} for a Gaussian prior).

For finite $\sigma$, the penalty increases with both the likelihood threshold
$\lambda$ (i.e.\ the distance from the minimum of the log-likelihood) and with the
uncertainty $\sigma$ on $\ln J$, quantifying the degradation of the constraint on $x$
due to imperfect knowledge of the astrophysical normalization.
Unlike the linear-Gaussian model in $J$, the log-normal case does not impose a sharp
condition such as $J_0^2>\lambda\sigma_J^2$; instead, the prior remains normalizable
for any finite $\sigma$, and the impact of large $\sigma$ is encoded smoothly through
$\mathcal{P}(\sigma)$.

\section{Validation with Monte Carlo simulations}
\label{sec:mc_validation}

To validate the analytical expressions derived in the previous sections, we
performed a series of Monte Carlo (MC) tests using a simplified toy model.
The goal of these tests is to verify that the penalty factors (see Eqs. \eqref{eq:gaussian_penalty} and  \eqref{eq:penalty_lognormal}),  which we recall being
\begin{equation}
    \mathcal{P}= \frac{1}{ \sqrt{1-\lambda r^2} }, \qquad  \mathcal{P}=\frac{\sqrt{\varepsilon}}{\sqrt{\lambda} \sigma}\,e^{\varepsilon},
\end{equation}
for the Gaussian and log-normal case, respectively, 
correctly reproduce the impact of
the $J$-factor uncertainty on the upper limits when the likelihood is evaluated
numerically.

Our MC simulations are based on a simplified ON/OFF counting experiment,
which retains the essential statistical ingredients of likelihoods commonly
used in gamma-ray analyses, namely Poisson-distributed event counts, a
background component treated as a nuisance parameter, and a signal model
whose normalization is proportional to the product $J\mu$. More detailed
features of experimental analyses, such as energy binning, spatial
templates, and instrumental response functions, are not required for the
validation presented here and are effectively absorbed into the signal
normalization.
For each pseudo-experiment we generate the observed counts in an OFF region,
$n_{\rm off}$, and in an ON region, $n_{\rm on}$, according to independent Poisson
processes,
\begin{equation}
n_{\rm off} \sim {\rm Pois}(b), \qquad
n_{\rm on} \sim {\rm Pois}(\alpha b + K\,J\,\mu),
\label{eq:onoff_model}
\end{equation}
where $b$ is the (unknown) background expectation in the OFF region, $\alpha$ is
the exposure ratio between ON and OFF, and $\mu\ge 0$ is the parameter controlling
the signal strength. The constant $K$ is a fixed normalization factor and $J$ is the astrophysical
$J$ factor; for the purpose of the toy simulations, both are set to unity, as
only relative effects enter the quantities of interest. In the following we  assume the background-only hypothesis, $\mu=0$, so that the true ON expectation is
$\alpha b$.

Given $(n_{\rm on},n_{\rm off})$, we build a profile-likelihood ratio test statistic
for a fixed signal-strength hypothesis $\mu$,
\begin{equation}
q (\mu ) \equiv
\begin{cases}
-2\ln\!\left[\dfrac{\mathcal{L}(\mu,\hat{\hat b}_\mu,\hat{\hat J}_\mu)}
{\mathcal{L}(\hat\mu,\hat b,\hat J)}\right],
& \hat\mu \le \mu,\\[1.2ex]
0, & \hat\mu > \mu,
\end{cases}
\label{eq:qmu_def}
\end{equation}
where hats denote unconditional maximum-likelihood estimators (MLEs), while double
hats denote conditional MLEs evaluated at fixed $\mu$. The likelihood entering
Eq.~\eqref{eq:qmu_def} is the product of the ON/OFF Poisson terms and (when
included) an external constraint on $J$,
\begin{equation}
\mathcal{L}(\mu,b,J) =
{\rm Pois}(n_{\rm on}|\alpha b + J\mu)\;
{\rm Pois}(n_{\rm off}|b)\;
\pi(J).
\label{eq:toy_likelihood}
\end{equation}
For the prior $\pi(J)$ we consider both a Gaussian function in
$J$ centered at $J_0$ with width $\sigma_J$, and  a log-normal prior, i.e. a
Gaussian prior in $\ln J$ centered at $ln J_0$ with width $\sigma$.

\begin{figure*}[h!t]
    \centering
    \includegraphics[width=0.45\linewidth]{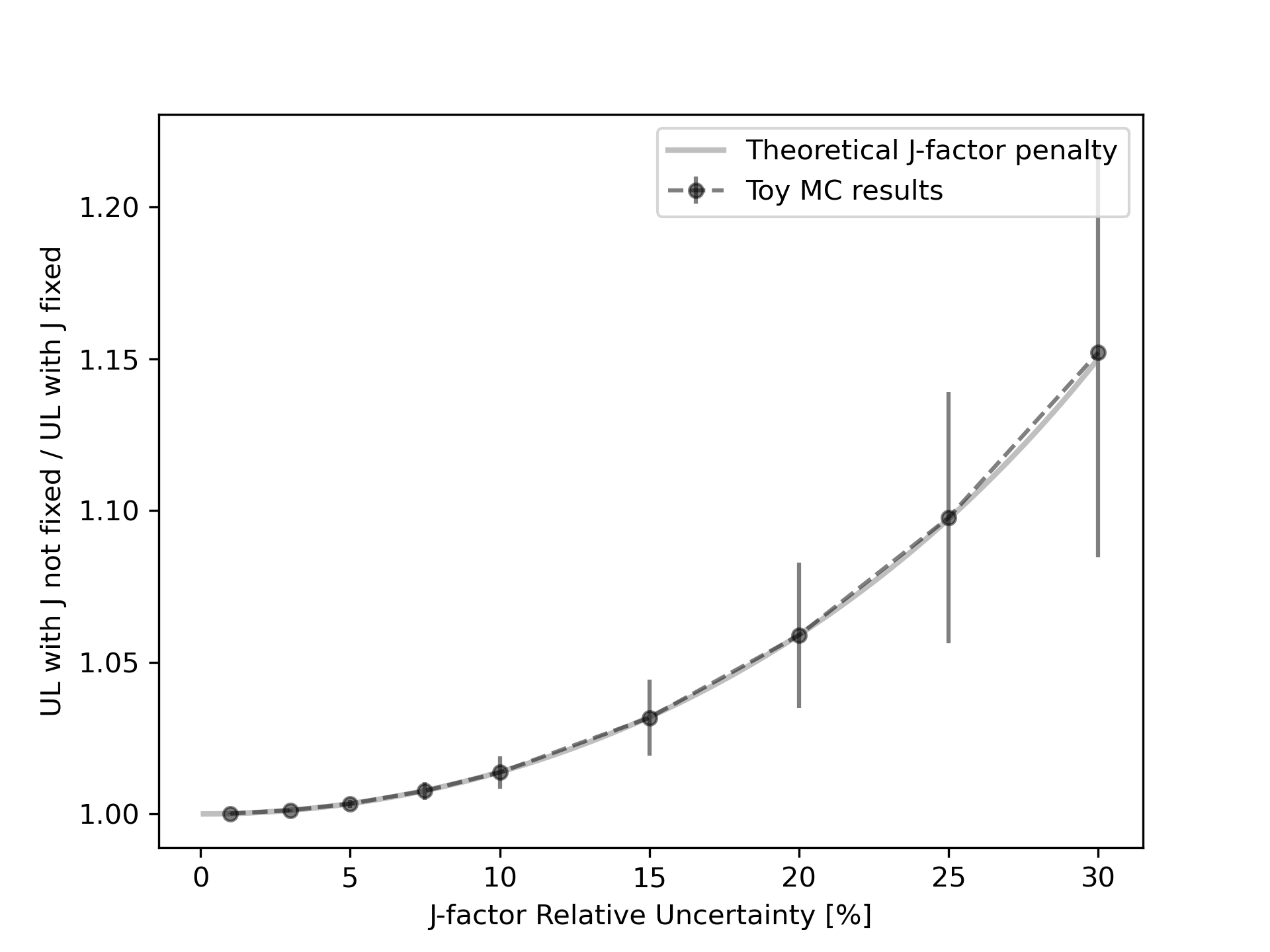}
      \includegraphics[width=0.45\linewidth]{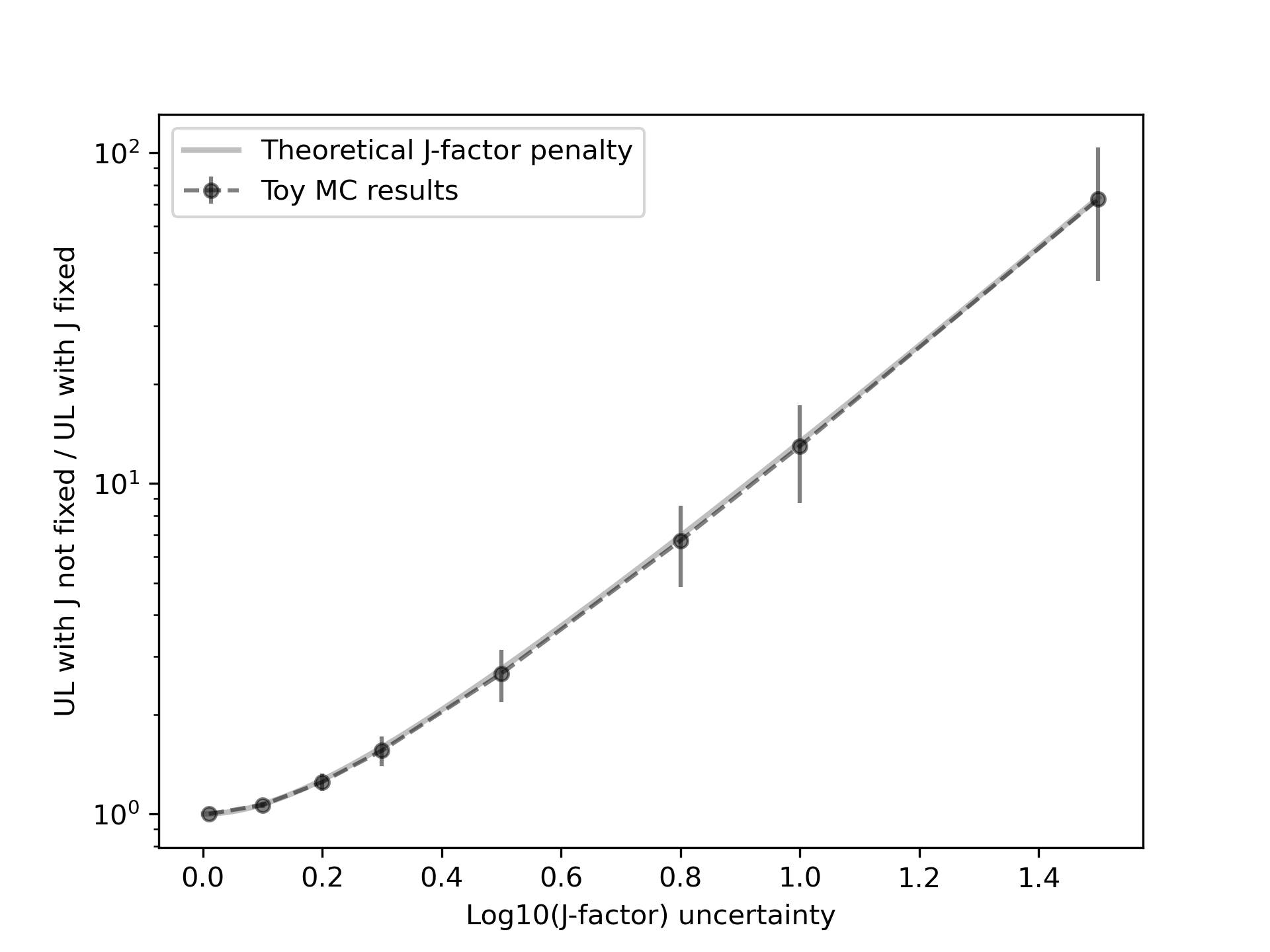}
    \caption{Comparison between the analytical $J$-factor penalty $\mathcal{P}$ and the
corresponding estimate obtained from toy Monte Carlo simulations,
$\mathcal{P}_{\rm MC}$, defined as the ratio of upper limits with profiled and
fixed $J$ factor.
Left: Gaussian prior on $J$, shown as a function of the relative
$J$-factor uncertainty $\sigma_J/J_0$.
Right: log-normal prior on $J$, shown as a function of the uncertainty on
$\log{10} J$.
Solid lines denote the analytical predictions from
Eqs.\eqref{eq:gaussian_penalty} and \eqref{eq:penalty_lognormal}, while markers
show the mean over toy MC realizations, with error bars representing the standard deviation. }
    \label{fig:mc_penalty}
\end{figure*}

For each toy dataset, the one-sided upper limit $\mu_{\rm UL}$ is obtained by
solving
\begin{equation}
q(\mu_{\rm UL}) = \lambda,
\label{eq:toy_ul_condition}
\end{equation}
with $\lambda=2.71$ for a one-sided $95\%$ CL using the usual asymptotic
prescription. We compute $\mu_{\rm UL}$ in two configurations: a fixed-$J$ analysis
with $J=J_0$, that we call $\mu_{\rm UL}^{\rm fix} $, and a profiled-$J$ analysis, that we call $\mu_{\rm UL}^{\rm prof}$,  in which $J$ is treated as a nuisance
parameter constrained by $\pi(J)$. The MC estimate of the penalty is then defined
as the ratio
\begin{equation}
\mathcal{P}_{\rm MC} \equiv
\frac{\mu_{\rm UL}^{\rm prof}}{\mu_{\rm UL}^{\rm fix}},
\label{eq:Pmc_def}
\end{equation}
which can be compared directly to the analytical expressions derived in
Sec.~\ref{sec:mc_validation}.

Figure~\ref{fig:mc_penalty}  shows the comparison
between $\mathcal{P}_{\rm MC}$ extracted from the MC simulations (points are the mean value, with error bars representing the standard deviation) and the
analytical penalty factors derived in  Eqs. \eqref{eq:gaussian_penalty} and  \eqref{eq:penalty_lognormal} for the Gaussian (left plot) and log-normal (right plot)
priors, respectively. In both cases, a good agreement is observed over
the full range of tested $J$-factor uncertainties. The MC points follow the
analytical predictions, confirming the validity of the $J$-factor penalty derived in this work.

These tests demonstrate that, within the validity of the quadratic approximation adopted in this work, the impact of astrophysical $J$-factor uncertainties on the derived upper limits is accurately described by the analytical penalty factors $\mathcal{P}(\sigma_J)$. Combined with the trivial rescaling associated with the central value of the $J$ factor, these penalty factors provide a practical prescription for updating published limits when revised astrophysical determinations become available.

\subsection{Stability under successive reinterpretations}
\label{sec:stability_chain}

One potential concern regarding the proposed framework is that repeated 
applications of the update prescription could lead to an uncontrolled accumulation 
of approximation errors. Such a situation may arise if improved determinations of 
the astrophysical $J$-factor become available over time, requiring previously 
published upper limits to be successively reinterpreted in the absence of the 
original experimental likelihood.

To quantify this effect, we performed an additional toy Monte Carlo study using 
the same ON/OFF setup described in Sec.~\ref{sec:mc_validation}. We considered a 
sequence of four log-normal $J$-factor priors with identical central values but 
progressively smaller uncertainties,
\begin{equation}
\sigma_{\log_{10}J} = 0.5,\; 0.4,\; 0.3,\; 0.2,
\end{equation}
mimicking three successive improvements in the astrophysical determination of the 
$J$ factor. This sequence is chosen to be representative of the realistic regime 
of upcoming $J$-factor revisions. Current dwarf-spheroidal $J$-factor 
uncertainties typically lie in the range $\sigma_{\log_{10}J} \sim 0.3\text{--}0.5$ 
(see, e.g., Refs.~\cite{pace2019scaling,abe2025prospects}), and forthcoming improvements 
from Gaia, DESI, WEAVE, and Euclid are expected to progressively reduce these 
uncertainties towards $\sigma_{\log_{10}J} \sim 0.2$ or below for the 
best-characterized targets. The sequence above, therefore, corresponds to a 
plausible scenario in which several successive astrophysical analyses 
progressively refine the $J$-factor determination of a given target.

For each value of $\sigma_{\log_{10}J}$, the upper limit was first obtained by 
profiling the likelihood over the nuisance parameter $J$. This represents the 
ideal situation in which the complete statistical analysis is repeated every time 
the astrophysical prior is updated.

We then considered the more realistic scenario in which the original analysis 
cannot be repeated from scratch with the updated $J$-factor uncertainties. 
Starting from the upper limit $\mu_{\rm prof}^{(0)}$ corresponding to the first 
prior, each subsequent limit was reconstructed by successively applying the 
analytical update prescription,
\begin{equation}
\mu_{\rm rec}^{(n+1)}
=
\mu_{\rm rec}^{(n)}
\frac{P(\sigma_{n+1})}{P(\sigma_n)},
\qquad
\mu_{\rm rec}^{(0)} \equiv \mu_{\rm prof}^{(0)},
\end{equation}
where $P(\sigma)$ is the log-normal penalty factor derived in 
Eq.~\ref{eq:penalty_lognormal}. The resulting limits were compared with those 
obtained by fully re-profiling the likelihood for each updated prior.

\begin{figure}[t]
\centering
\includegraphics[width=0.95\linewidth]{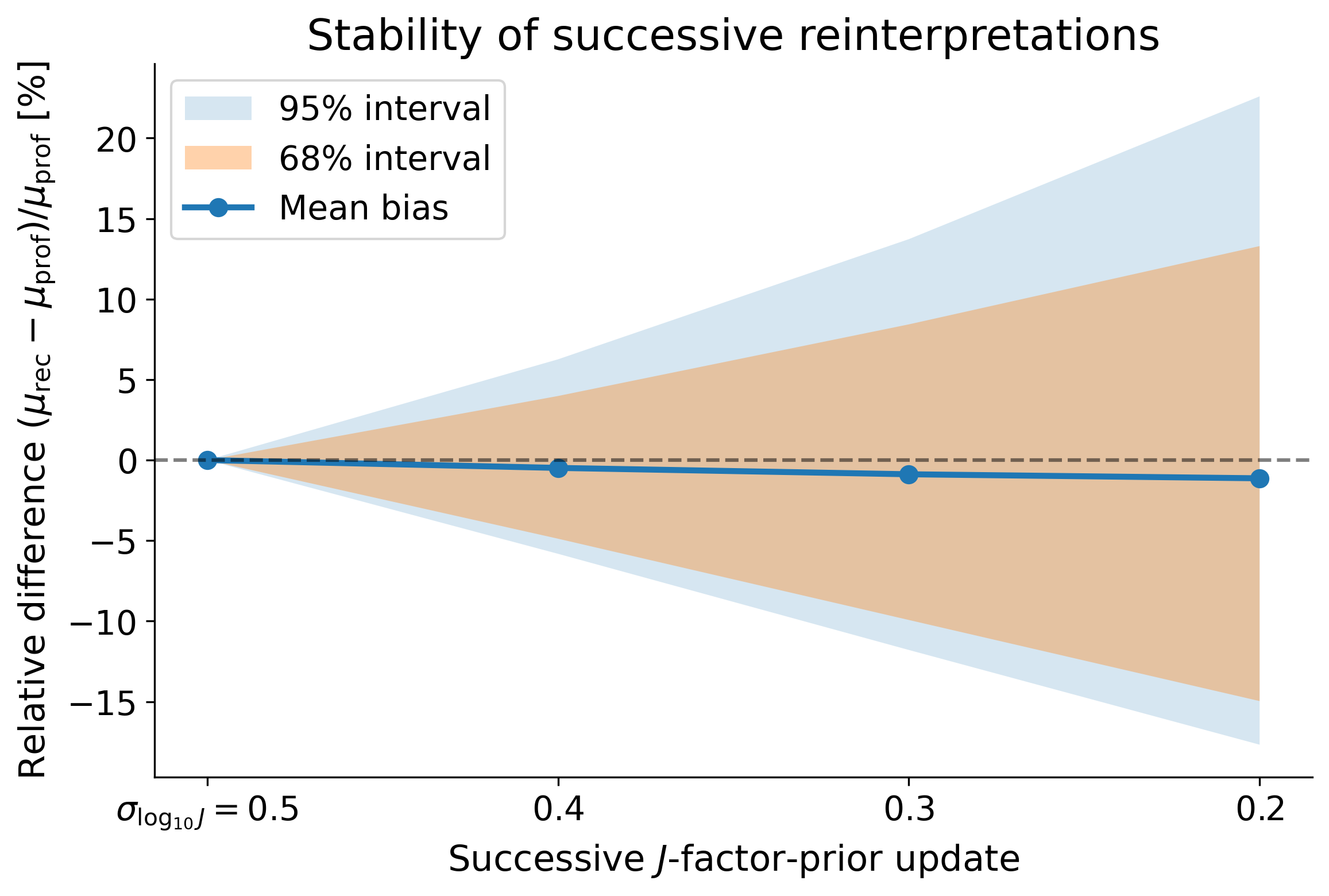}
\caption{Mean relative difference 
between upper limits obtained by successively applying the analytical update 
prescription, $\mu_{\rm rec}$, and those obtained by re-profiling the likelihood 
for each updated $J$-factor prior, $\mu_{\rm prof}$. The four points correspond 
to a sequence of log-normal $J$-factor uncertainties, 
$\sigma_{\log_{10}J} = 0.5,\, 0.4,\, 0.3,\, 0.2$, representative of the 
progressive refinement of $J$-factor determinations expected from current and 
forthcoming astrophysical programs. The shaded bands represent the central 
$68\%$ and $95\%$ intervals obtained from 500 toy Monte Carlo realizations.}
\label{fig:chain}
\end{figure}

Figure~\ref{fig:chain} summarizes the results of this test. The solid blue line 
shows the mean relative difference between the successively reinterpreted limits 
and those obtained from a complete reanalysis of the toy likelihood, while the 
shaded regions indicate the central $68\%$ and $95\%$ intervals of the 
distribution. Averaged over 500 toy Monte Carlo realizations, the mean relative 
differences are found to be
\begin{equation}
\left\langle
\frac{\mu_{\rm rec}-\mu_{\rm prof}}
{\mu_{\rm prof}}
\right\rangle
=
\left(
0,\;
-0.49,\;
-0.88,\;
-1.13
\right)\%,
\end{equation}
corresponding to the four values of $\sigma_{\log_{10}J}$ listed above. The first 
entry is zero by construction, since these upper limits, being the first of the 
chain, do not require any reinterpretation.

The test demonstrates that, in the regime of realistic $J$-factor revisions, the 
cumulative mean bias remains at the $\sim 1\%$ level even after three consecutive 
reinterpretations, with the distributions remaining centred close to zero. The 
per-realization spread is also moderate: after a single update the $68\%$ 
interval is contained within $\pm 5\%$, while after three successive updates it 
stays within roughly $\pm 13\%$, with the $95\%$ interval remaining within 
$\pm 20\%$. The width of the distribution grows only modestly with the number of 
successive updates and is small compared with the typical $J$-factor uncertainty 
itself, which translates into a comparable or larger uncertainty on the 
published limits.

As a  conservative practice, updated limits should always be obtained by applying the 
prescription directly to the original published upper limit and the original 
$J$-factor prior adopted in the experimental analysis, rather than by repeatedly 
reinterpreting already updated results. This approach avoids the unnecessary 
propagation of approximation errors and ensures that each reinterpretation remains 
as close as possible to the original statistical inference. Naturally, whenever 
the original experimental likelihood is available, a complete reanalysis remains 
the preferred approach.

\section{Reproducing published limits with updated $J$-factor uncertainties}
\label{sec:recast_realdata}

As a final validation of the method, we apply the analytical penalty formalism
to a published set of experimental limits and verify that it correctly
reproduces results obtained with a full treatment of $J$-factor uncertainties.
For this purpose, we consider Ref.~\cite{abe2025prospects}, which presents
projected upper limits on the dark matter annihilation cross section from
gamma-ray observations of multiple targets, explicitly accounting for
astrophysical uncertainties.

In particular, we focus on Fig.~8 of Ref.~\cite{abe2025prospects}, where upper
limits are shown including the effect of $J$-factor uncertainties.
The same work also reports limits obtained under the assumption of fixed
$J$ factors, as well as the corresponding $J$-factor uncertainties for each
source assuming a log-normal distribution (see Table~A5 of Ref.~\cite{abe2025prospects}).

Starting from the limits computed with fixed $J$, we independently reconstruct
the corresponding limits including astrophysical uncertainties by applying the
penalty prescription derived in Sec.\ref{sec:Penalty}. For each target, we compute the
appropriate penalty factor $\mathcal{P}$ using the reported $J$-factor
uncertainty and multiply the fixed-$J$ upper limit by this factor. No additional
information from the gamma-ray likelihood or instrumental response is used.

\begin{figure*}[h!t]
    \centering
    \includegraphics[width=0.95\linewidth]{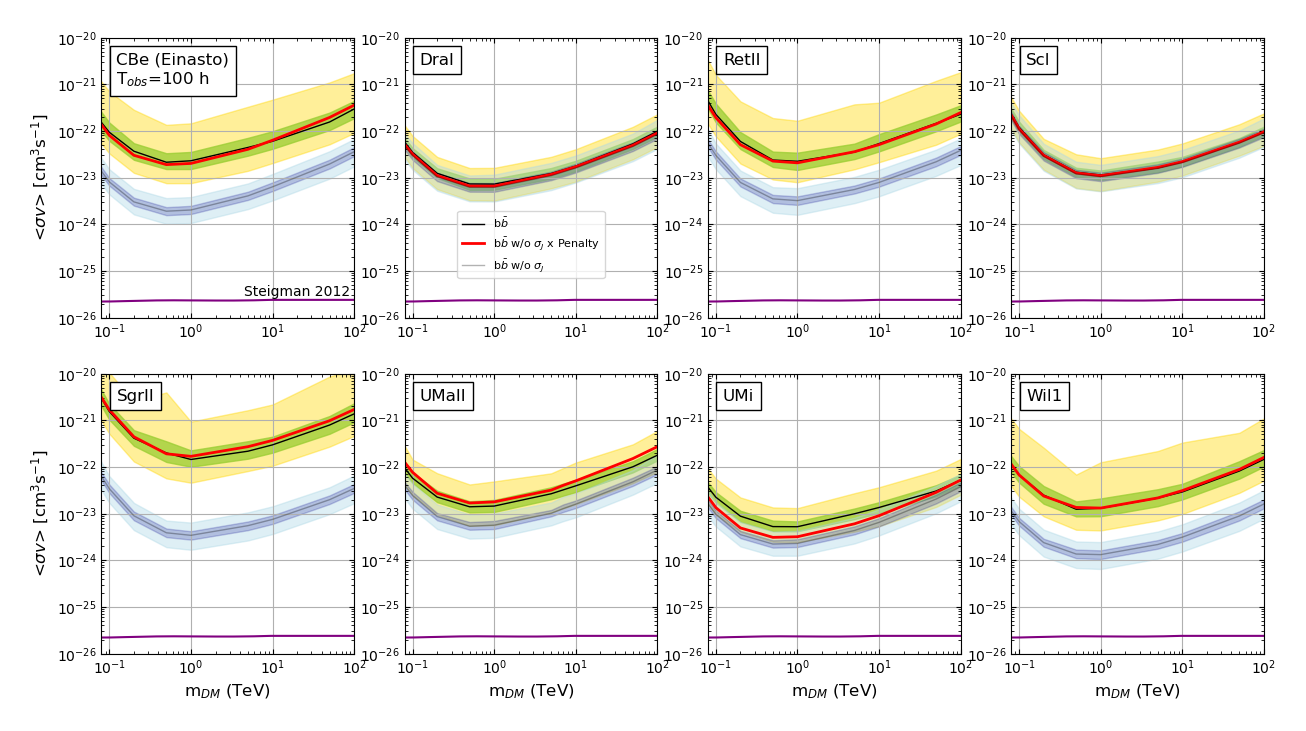}
    \caption{ Upper limits on the dark matter annihilation cross section
$\langle\sigma v\rangle$ as a function of the dark matter mass
$m_{\rm DM}$ for the $b\bar b$ annihilation channel, for the set of
targets considered in Ref.~\cite{abe2025prospects}.
The black curve, together with the green ($1\sigma$) and yellow ($2\sigma$)
bands, reproduces the published median expected limit including the
$J$-factor uncertainty. The blue curve, together with the blue ($1\sigma$)
and light-blue ($2\sigma$) bands, shows the corresponding expected limit
obtained when the $J$-factor uncertainty is excluded. The red curve is
obtained by multiplying the latter by the analytical $J$-factor penalty
derived in Sec.~\ref{sec:Penalty}, using the $J$-factor uncertainties
reported in Table A5 of Ref.~\cite{abe2025prospects}.}
    \label{fig:abe2025_recast}
\end{figure*}

Figure~\ref{fig:abe2025_recast} shows the result of this procedure. The original
limits from Ref.~\cite{abe2025prospects} are shown\footnote{The reproduction of the original limits and the direct comparison
presented here were made possible by the public repository released by the
authors of Ref.~\cite{abe2025prospects}, which provides Jupyter notebooks to
reproduce most of the figures in the paper and is available at
\url{https://zenodo.org/records/17079907}.} together with the limits
reconstructed using the penalty method.

The two sets of limits are found to be in very good agreement over the full mass
range for all considered targets. For most sources, the limits reconstructed
using the penalty prescription lie well within the quoted $1\sigma$ uncertainty
bands of Ref.~\cite{abe2025prospects}. A larger deviation is observed for Ursa Minor (UMi), where the reconstructed limits remain compatible with the
published results at the $\sim 2\sigma$ level. This difference may arise from
source-specific aspects of the likelihood construction or from a different
implementation of the $J$-factor uncertainty treatment in the original analysis.

\section{Combination of multiple targets}
\label{sec:combined_likelihood}

The framework developed above can be straightforwardly generalized to the case
in which the total likelihood is obtained by combining independent likelihoods
from multiple targets, each characterized by its own astrophysical $J$ factor
and associated uncertainty. This situation commonly arises in combined analyses
of several dwarf spheroidal galaxies or, more generally, when upper limits from
different sources are combined at the likelihood level.

Assuming statistical independence, the total log-likelihood is given by the sum
of the individual contributions. In the quadratic approximation and adopting a
log-normal parametrization for the $J$-factor uncertainties, the combined
log-likelihood can be written as
\begin{equation}
-2\ln \mathcal{L}(x,\{y_i\})
\simeq
\sum_i
\left[
A_i \left(e^{y_i}x-\hat z\right)^2
+
B_i (y_i-y_{0,i})^2
\right],
\label{eq:combined_ll_start}
\end{equation}
where $x\equiv\langle\sigma v\rangle$ is the common annihilation cross section,
$y_i\equiv\ln J_i$, $y_{0,i}=\ln J_{0,i}$, $B_i=1/\sigma_i^2$, and $A_i$ encodes the
curvature of the likelihood for the $i$-th source. Since we are interested in setting upper limits in the absence of a signal, we
work under the null-hypothesis assumption $\hat z=0$.

Because the nuisance parameters $y_i$ appear only in the $i$-th term of the sum
in Eq.~\eqref{eq:combined_ll_start}, profiling over $\{y_i\}$ factorizes target
by target for any fixed $x$,
\begin{equation}
0= \frac{\partial}{\partial y_i}
\left[
-2\ln \mathcal{L}(x,\{y_i\})
\right]= 2 A_i e^{2y_i} x^2 + 2 B_i (y_i-y_{0,i})
\label{eq:profile_factorization}
\end{equation}
Defining $\varepsilon_i\equiv y_{0,i}-y_i\ge 0$ and using $J_{0,i}=e^{y_{0,i}}$,
the Eq. \eqref{eq:profile_factorization} becomes
\begin{equation}
A_i J_{0,i}^2 x^2 e^{-2\varepsilon_i} = B_i \varepsilon_i.
\label{eq:eps_stationary}
\end{equation}
This implicit equation admits a closed-form solution for $\varepsilon_i(x)$ in
terms of the Lambert-$W$ function,
\begin{equation}
\varepsilon_i(x)
=
\frac{1}{2}\,
W\!\left( 2\, A_i \, J_{0,i}^2 \, \sigma_i^2  \; \, x^2  \right),
\label{eq:eps_lambert}
\end{equation}
where $W$ denotes the principal branch~\footnote{The Lambert-$W$ function is defined as the solution of
$W(z)\,e^{W(z)}=z$. For real and non-negative arguments, the principal branch
$W_0$ is real and single-valued. The argument in
Eq.~\eqref{eq:eps_lambert} is non-negative for $x\ge 0$, ensuring that the
principal branch yields a real solution.}.

Evaluating Eq.~\eqref{eq:combined_ll_start} at the profiled values $y_i(x)$ and
using Eq.~\eqref{eq:eps_stationary} to eliminate the term
$A_i J_{0,i}^2 x^2 e^{-2\varepsilon_i}$, from Eq. \eqref{eq:combined_ll_start} the profiled log-likelihood is 
\begin{equation}
-2\ln \mathcal{L}(x)
=
\sum_i B_i\left[\varepsilon_i(x) + \varepsilon_i^2(x)\right]
\equiv \sum_i Q_i(x),
\label{eq:qi_def}
\end{equation}
where we have defined 
\begin{equation}
Q_i(x)\equiv B_i\,\varepsilon_i(x)\bigl[1+\varepsilon_i(x)\bigr]
\label{eq:qtot_def}
\end{equation}
with $\varepsilon_i(x)$ given by Eq.~\eqref{eq:eps_lambert}.

\subsection{Combined upper limits}
\label{sec:combibed_uls}

The combined upper limit $x_{\rm UL}$ is obtained by solving the
standard threshold condition
\begin{equation}
-2\ln \mathcal{L}(x_{\rm UL}) = \sum_{i=1}^{N_{\rm tgt}} Q_i(x_{\rm UL}) =  \lambda,
\label{eq:comb_ul_condition}
\end{equation}
with $\lambda=2.71$ for a one-sided $95\%$ CL.

Unlike the single-target case, for which a solution is given in Eq. \eqref{eq:xul_lognormal_general}, the multi-target case in  Eq.~\eqref{eq:comb_ul_condition} does not yield a
closed-form expression for $x_{\rm UL}$ in general, because it involves a sum of
nonlinear functions of $x$.

However, the solution is straightforward to obtain numerically: for $x\ge 0$,
each $Q_i(x)$ is monotonic increasing (as expected for a profile likelihood), and therefore the left-hand side of
Eq.~\eqref{eq:comb_ul_condition} is a one-dimensional monotonic function. As a
result, standard bracketing and root-finding algorithms (e.g.\ Brent's method)
converge rapidly and robustly.

In practical applications, when combining published upper limits from different
targets, all the ingredients required to construct the functions $Q_i$ defined
in Eq.~\eqref{eq:qi_def} are available from public information. Specifically,
one needs:
\begin{itemize}
    \item the central value and uncertainty of the astrophysical $J$ factor for
    each target, $J_{0,i}$ and $\sigma_i$, which enter the definition of
    $\varepsilon_i(x)$ in Eq.~\eqref{eq:eps_lambert};
    \item the coefficients $A_i$ appearing in Eq.~\eqref{eq:eps_lambert}, which
    can be reconstructed from the single-target upper limit
    $x_{{\rm UL},i}$ using the single-target log-normal result
    Eq.~\eqref{eq:xul_lognormal_general},
\end{itemize}
\begin{equation}
A_i
=
\frac{1}{x_{{\rm UL},i}^2}\,
\frac{1}{J_{0,i}^2}
\left(
\frac{\sqrt{\varepsilon_{i,{\rm UL}}}}{\sigma_i}\,e^{\varepsilon_{i,{\rm UL}}}
\right)^2,
\end{equation}
where
\begin{equation}
\varepsilon_{i,{\rm UL}}
=
\frac{1}{2}\left(\sqrt{1+4\lambda\sigma_i^2}-1\right).
\end{equation}
We emphasize that, when combining upper limits using updated determinations of
the $J$ factors, the values of $J_{0,i}$ and $\sigma_i$ entering the
reconstruction of $A_i$ must correspond to the \emph{original} $J$-factor
assumptions under which the single-target limits were derived, and therefore
need not coincide with the updated values used in the definition of
$\varepsilon_i(x)$.
At this point, once the $A_i$ are reconstructed at each dark-matter mass, one can combine ULs  by solving
Eq.~\eqref{eq:comb_ul_condition} numerically, without requiring access to the
full experimental likelihood.

\subsubsection{Validation with published combined limits}
\label{sec:validation_magic}

As a validation of the multi-target numerical procedure described in
Sec.~\ref{sec:combibed_uls}, we tested the method against published combined upper
limits from the MAGIC Collaboration.
In particular, we considered the limits reported in Fig.~4 of
Ref.~\cite{MAGIC:2021mog} for the $\tau^+\tau^-$ annihilation channel, which are
obtained by combining observations of several dwarf spheroidal galaxies.

The single-target upper limits entering the combination were taken directly from
the published MAGIC results. Specifically, the limits used in this test are currently available in the
\texttt{gDMbounds} repository, which collects up-to-date indirect dark matter
constraints in machine-readable form\footnote{The database is publicly
available at \url{https://github.com/micheledoro/gDMbounds/}.}.
For each target, we used the corresponding $J$-factor central value and
uncertainty quoted in Ref.~\cite{MAGIC:2021mog}.

Following the procedure outlined in Sec.~\ref{sec:combibed_uls}, we reconstructed
the coefficients $A_i$ from the single-target limits and numerically solved the
combined profile-likelihood condition,
Eq.~\eqref{eq:comb_ul_condition}, to obtain the combined upper limit as a function
of the dark matter mass.
No information beyond the published upper limits and $J$-factor uncertainties
was required.

\begin{figure}[h]
    \centering
    \includegraphics[width=0.9\linewidth]{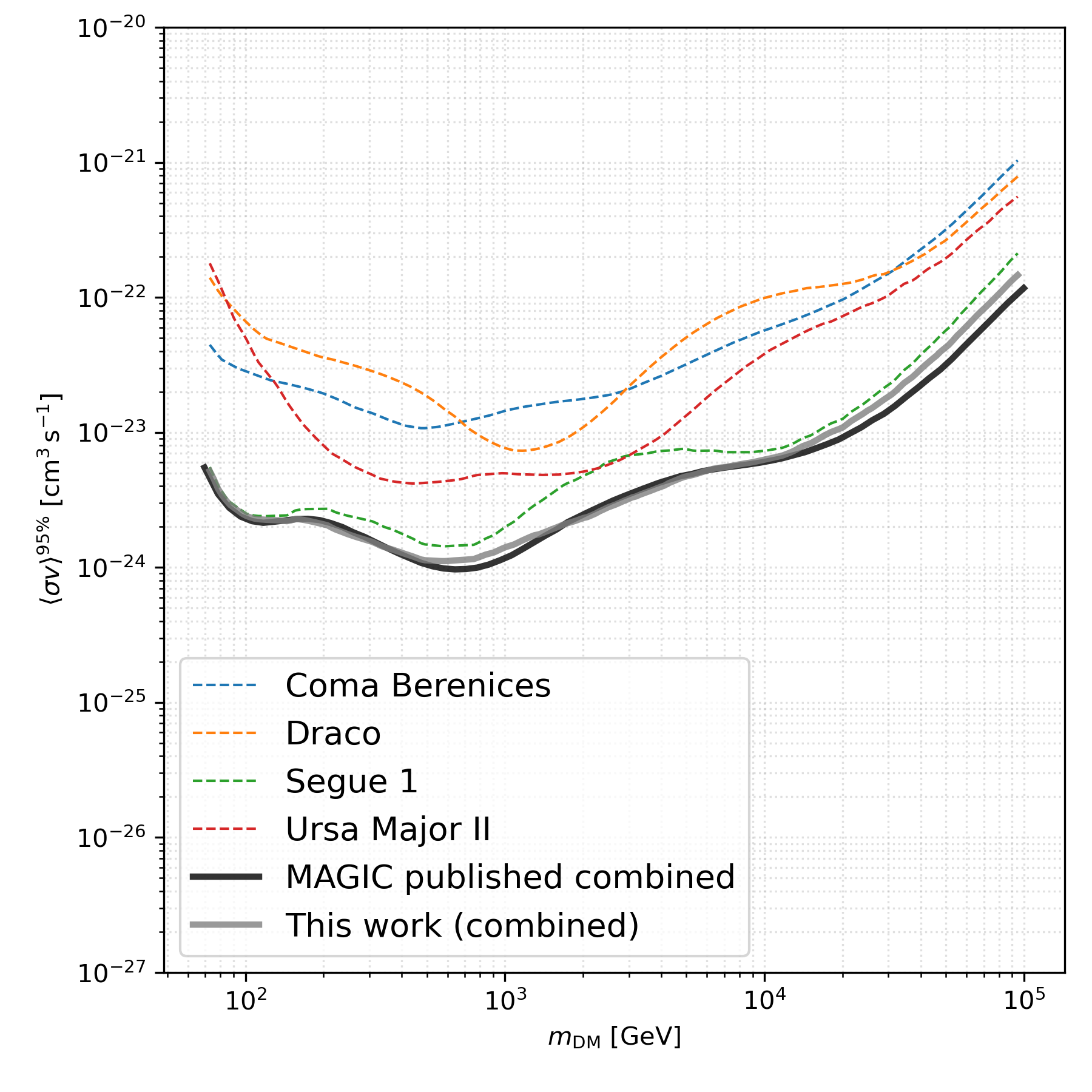}
    \caption{
    Upper limits at $95\%$ CL on the dark matter annihilation cross section
    $\langle\sigma v\rangle$ into the $\tau^+\tau^-$ channel as a function of the
    dark matter mass.
    The dashed colored curves show the single-target limits published by the MAGIC
    Collaboration~\cite{MAGIC:2021mog} for the individual
    dwarf spheroidal galaxies entering the combination.
    The black solid line corresponds to the combined limit published by the MAGIC
    Collaboration~\cite{MAGIC:2021mog} for the same annihilation channel, while the gray solid line shows the
    combined upper limit reconstructed in this work using the numerical
    profile-likelihood combination described in Sec.~\ref{sec:combibed_uls}.}
    \label{fig:magic_validation}
\end{figure}

Figure~\ref{fig:magic_validation} shows the result of this comparison.
The dashed colored curves correspond to the single-target upper limits for the
individual dwarf spheroidal galaxies entering the combination.
The black solid line reproduces the combined upper limit published by the MAGIC
Collaboration~\cite{MAGIC:2021mog}, while the gray solid line shows the combined limit obtained with
the numerical procedure described in this work.

Overall, good agreement is observed between the published combined limit and
the reconstructed one over the full dark matter mass range considered.
Small residual differences appear only at the highest masses.
These deviations are plausibly attributable to subleading effects beyond the
quadratic approximation, as well as to mild departures from the strict
null-hypothesis assumption due to small positive or negative
fluctuations in the data that are not statistically significant.
Importantly, the observed deviations remain well within the $1\sigma$
uncertainty bands reported in Ref. ~\cite{MAGIC:2021mog}.

\section{Conclusions}
\label{sec:conclusions}

In this work we have presented a general and practical framework to update
published upper limits on the dark matter annihilation cross section when revised
determinations of the astrophysical $J$ factor become available. We derived analytical expressions that quantify
the impact of $J$-factor uncertainties for single targets, both for Gaussian and
log-normal priors, and we validated these results with both toy Monte Carlo simulations and published limits that explicitly include
astrophysical uncertainties. When combined with previously developed recasting approaches \cite{d2025recasting} that focus on changes in the dark matter particle-physics model, this method allows existing limits to be consistently updated to reflect both advances in particle-physics modeling and refined astrophysical determinations, without requiring access to the full experimental likelihood.

We further showed that, while a closed-form analytical expression does not exist
in general for the combination of multiple targets, the formalism naturally
extends to this case through a simple and robust numerical procedure.
The required inputs can be reconstructed directly from published single-target
limits and quoted $J$-factor uncertainties, without access to the full
experimental likelihood.
A validation against published combined limits from the MAGIC Collaboration
demonstrates that this approach accurately captures the effect of astrophysical
uncertainties in realistic multi-target analyses.

The central motivation of this work is the fact that current indirect-detection constraints are intrinsically tied to the astrophysical assumptions adopted at the time of the analysis, while significant progress in the determination of dark-matter distributions is expected in the coming years. The framework presented here provides a practical and statistically motivated procedure to approximately reinterpret existing limits as improved $J$-factor determinations become available, thereby extending the scientific lifetime of published indirect-detection results in situations where a full reanalysis of the original experimental likelihood is not feasible. The proposed prescription is therefore intended as a complementary reinterpretation tool rather than as a replacement for official experimental analyses. Whenever the original likelihood is available, a complete reanalysis remains the preferred procedure.

Finally, we briefly comment on the case of decaying dark matter.
The formalism developed for annihilating dark matter applies almost unchanged to
this scenario.
The main differences are that the kinematics involve the replacement
$m_\chi \rightarrow m_\chi/2$, and that the parameter of interest is the inverse
lifetime $x \equiv 1/\tau$ rather than the annihilation cross section
$\langle\sigma v\rangle$.
Since all kinematic information is absorbed into the coefficient $A$, this
substitution does not affect the derivation of the astrophysical penalty factors.
As a consequence, starting from the generic upper-limit condition
Eq.~\eqref{eq:generic_ul_form}, the resulting constraint on the dark matter
lifetime takes the form of a \emph{lower} limit,
\begin{equation}
\tau_{\rm LL}
=
\frac{\sqrt{A}}{\sqrt{\lambda}}\,
J_0\,
\mathcal{P}_{\rm decay}(\sigma),
\end{equation}
where $J_0$ now denotes the astrophysical factor relevant for decay and
$\mathcal{P}_{\rm decay}(\sigma)$ is simply 
$1/\mathcal{P}(\sigma)$ with $\mathcal{P}$ the the penalty factor obtained in the annihilation case.

In summary, the methods presented here provide a simple and computationally inexpensive tool to reinterpret existing indirect-detection limits in light of future improvements in astrophysical modeling, without requiring access to the full experimental likelihood. While developed in the context of dark matter searches affected by $J$-factor uncertainties, the underlying framework is more general and applies to a broad class of problems in which the predicted signal strength depends multiplicatively on external nuisance parameters.  As experimental data continue to outlive specific modeling assumptions, such reinterpretation frameworks will become increasingly important for maximizing the scientific return of indirect detection experiments.

\begin{acknowledgments}
We thank Michele Doro for carefully reading the manuscript and providing valuable comments and suggestions that helped improve this work. We also thank Francesco Gabriele Saturni and Gonzalo Rodriguez-Fernandez for useful input when validating the method with the published limits in Ref.~\cite{abe2025prospects}. We thank the anonymous referee for their careful assessment of this work and for the constructive comments and suggestions, which helped improve the manuscript.
\end{acknowledgments}

\paragraph*{Data Availability}
Results presented in this work can be reproduced from scratch using the
publicly available code at \hyperlink{https://github.com/giacomodamico24/Revise-DM-limits-with-Updated-JFactor-priors/}{github.com/giacomodamico24/Revise-DM-limits-with-Updated-JFactor-priors}.

\paragraph*{Funding}
The work of GDA on this project was supported by the Beatriu de Pin\'{o}s program (BP 2023).

\bibliographystyle{apsrev4-2}
\bibliography{references}

\end{document}